# The Murchison Widefield Array: Design Overview

Colin J. Lonsdale, Roger J. Cappallo, Miguel F. Morales, Frank H. Briggs,
Leonid Benkevitch, Judd D. Bowman, John D. Bunton, Steven Burns, Brian E. Corey, Ludi deSouza,
Sheperd S. Doeleman, Mark Derome, Avinash Deshpande, M. R. Gopalakrishna, Lincoln J. Greenhill,
David Herne, Jacqueline N. Hewitt, P. A. Kamini, Justin C. Kasper, Barton B. Kincaid,
Jonathan Kocz, Errol Kowald, Eric Kratzenberg, Deepak Kumar, Mervyn J. Lynch, S. Madhavi,
Michael Matejek, Daniel Mitchell, Edward Morgan, Divya Oberoi, Steven Ord, Joseph
Pathikulangara, T. Prabu, Alan E.E. Rogers, Anish Roshi, Joseph E. Salah, Robert J. Sault, N. Udaya
Shankar, K. S. Srivani, Jamie Stevens, Steven Tingay, Annino Vaccarella, Mark Waterson,
Randall B. Wayth, Rachel L. Webster, Alan R. Whitney, Andrew Williams, Christopher Williams.

*Abstract*— The Murchison Widefield Array (MWA) is a dipole-based aperture array synthesis telescope designed to operate in the 80-300 MHz frequency range. It is capable of a wide range of science investigations, but is initially focused on three key science projects. These are detection and characterization of 3-dimensional brightness temperature fluctuations in the 21cm line of neutral hydrogen during the Epoch of Reionization (EoR) at redshifts from 6 to 10, solar imaging and remote sensing of the inner heliosphere via propagation effects on signals from distant background sources, and high-sensitivity exploration of the variable radio sky. The array design features 8192 dual-polarization broad-band active dipoles, arranged into 512 "tiles" comprising 16 dipoles each. The tiles are quasi-randomly distributed over an aperture 1.5km in diameter, with a small number of outliers extending to 3km. All tile-tile baselines are correlated in custom FPGA-based hardware, yielding a Nyquist-sampled instantaneous monochromatic uv coverage and unprecedented point spread function (PSF) quality. The correlated data are calibrated in real time using novel position-dependent self-calibration algorithms. The array is located in the Murchison region of outback Western Australia. This region is characterized by extremely low population density and a superbly radio-quiet environment, allowing full exploitation of the instrumental capabilities.

*Index Terms*—Antenna arrays, Astronomy, Calibration, Imaging, Ionosphere.

Manuscript received November 14, 2008. This work was supported by the U.S. National Science Foundation under Grant AST-0457585. The Australian Research Council has contributed through grants LE775621, LE882938, and DP345001. Funding support was also provided by the U.S. Air Force Office of Scientific Research under Grant FA9550-0510247.

C. J. Lonsdale is with MIT Haystack Observatory, Westford, MA 01886 USA (phone: 781-981-5542; fax: 781-981-0590; e-mail: cjl@haystack.mit.edu). L. Benkevitch, R. J. Cappallo, B. Corey, S. Doeleman, M. Derome, B. Kincaid, E. Kratzenberg, D. Oberoi, A. E. E. Rogers, J. E. Salah, and A. R. Whitney are with MIT Haystack Observatory, Westford, MA USA.

M. Morales, formerly at MIT, is now with the University of Washington, Seattle, WA USA.

F. H. Briggs, J. Kocz, E. Kowald, A. Vaccarella and M. Waterson are with the Australian National University, Canberra, Australia.

J. D. Bowman, formerly at MIT, is now with the California Institute of Technology, Pasadena, CA USA

J. D. Bunton, L. deSouza and J. Pathikulangara are with Commonwealth Scientific and Industrial Research Organization, Epping, NSW, Australia.

S. Burns is with Burns Industries, Inc. Nashua, NH USA.

A. Deshpande, K. Deepak, M. R. Gopalakrishna, P. A. Kamini, S. Madhavi, D. Prabu, A. Roshi, N. U. Shankar and K. S. Srivani, are with the Raman Research Institute, Bangalore, India.

L. Greenhill, J. C. Kasper, D. Mitchell, S. Ord, and R. B. Wayth are with the Harvard-Smithsonian Center for Astrophysics, Cambridge, MA USA.

D. Herne, M. J. Lynch and S. Tingay are with Curtin University of Technology, Perth, Western Australia.

J. N. Hewitt, E. Morgan, M. Matejek and C. Williams are with the MIT Kavli Institute for Astrophysics and Space Research, Cambridge, MA USA

R. Sault and R. L. Webster are with the University of Melbourne, Australia.

J. Stevens is with the University of Tasmania, Tasmania, Australia.

A. Williams is with the University of Western Australia, Perth, Australia.

## I. INTRODUCTION

In the early days of astronomical radio interferometry, it was recognized that, in order to generate radio images of the sky at high resolution, it was necessary to sample many spatial frequencies via a wide range of interferometer spacings and orientations. This led to the innovation of earth-rotation aperture synthesis [1], and a rapid progression of ever more capable imaging instruments operating at centimeter and meter wavelengths. This progression culminated in the ATCA, GMRT, MERLIN, VLA, VLBI networks, and Westerbork, all of which remain productive facilities to the current time. Of these instruments, the GMRT and VLA have been unique in the comprehensiveness of their antenna distributions and a resulting instantaneous uv coverage capable of meaningful and useful "snapshot" imaging.

Recent telescope design efforts have explored so-called "large-N" architectures, which employ hundreds or thousands of independent small antenna elements to comprehensively sample the Fourier plane and provide high-fidelity "radio-camera" imaging properties ([2], [3], [4]). An additional benefit of such designs is that small elements have wide fields of view, and are thus able to gather useful astronomical imaging information from a larger part of the sky per unit time than an array of equal collecting area composed of larger antennas. This is easily seen by consideration of the average sky brightness. For a given receiver temperature, the SNR (and imaging information content) of the data stream generated by the received astronomical signal waveform is independent of antenna size. The larger number of smaller antennas in a large-N design of given total collecting area thus results in the collection of more aggregate astronomically-useful imaging information, in inverse proportion to the ratio of antenna areas and in direct proportion to the solid angle covered by the primary beam of the antennas.

Taking this concept to its logical conclusion, for a given terrestrial collecting area, information content for astronomical imaging is theoretically maximized by covering that area of ground with antennas sensitive to the entire 2-pi steradians of the sky, and capturing their signals independently. The number of antennas required is minimized if their inter-element spacings are non-redundant, implying that they are packed no more densely than is required for Nyquist sampling of the incident electric field on the ground. The amount of imaging information gathered is unchanged if the antennas are spread out over a larger area – array angular resolution is increased at the cost of decreased surface brightness sensitivity and decreased point spread function (PSF) quality. The principal difficulty with this otherwise ideal approach is that handling the resulting data flow has hitherto been prohibitively expensive. For minimal 2-bit dual-polarization sampling (noise-dominated passband case), the data rate generated by the antennas is approximately $32\lambda^{-2}AB$ bits/sec where A is the collecting area and B is the bandwidth in Hz. For the example of a major radio facility with $10^4$ m$^2$ of collecting area, operating at 10cm with a processed bandwidth of 1 GHz, this yields a total data rate of 32 petabits/sec, which is currently unmanageable.

Nevertheless, at low radio frequencies and allowing some restriction in the field of view of the antennas, it is possible to significantly exploit the theoretical advantages of the above-described approach using current technology. This is the foundation of the MWA system design concept, in which the wavelength is ~2 meters, the processed bandwidth is ~30 MHz, the collecting area is ~$10^4$ m$^2$, and the antennas have a field of view of order 0.4sr, yielding a data rate of only ~120 Gbits/sec. (in practice we use 5 bits to handle low-frequency dynamic range constraints, yielding a data rate of ~300 Gbits/sec). The MWA system design is aimed at maximizing the total information content of the data generated by observations, thereby optimally constraining instrumental calibration parameters and the sky brightness distribution. In this paper we describe the system design in these terms and how it translates into specific hardware and software solutions.

It is important to recognize that the MWA system design represents a significant departure from prior practice, and the only practical way to measure the advantages and limits of this new approach is to build and operate a complete system of sufficient scope. For this reason, the MWA is best thought of as a demonstrator instrument.

## II. SCIENCE OBJECTIVES

The science goals that have driven the MWA design are dominated by three key projects. These are detection and characterization of the redshifted 21cm neutral hydrogen signals from the Epoch of Reionization (EoR), solar and heliospheric plasma studies, and monitoring of the radio sky for transients. Since the main purpose of this paper is to describe the MWA design, these science objectives are only briefly summarized below.

EOR refers to the period in the history of the universe when the largely neutral hydrogen began to get ionized by the radiation from the first of the luminous sources. Current theoretical expectations place this period in the 6-20 redshift range. The collecting area of the MWA will probably fall short of the sensitivity requirements for direct imaging of the sites of reionization, and the array is therefore optimized for statistical detection approaches [5]. The wide field of view provides a very cost effective way of increasing the sensitivity to the EOR power spectrum fluctuations by increasing the area of the sky that is sampled. The unprecedented u-v coverage provided by the large number of elements and a full cross-correlation architecture lead to an exceptionally good PSF for estimating and removing foreground contamination.

The high quality monochromatic snapshot capability of the MWA is ideally suited for high fidelity solar imaging studies aimed at locating and characterizing solar radio bursts. The MWA frequency range is also well suited to studying the heliosphere using radio propagation effects, and the wide fields of view provide access to a large part of the heliosphere at any given time. The MWA will exploit these advantages in two ways. The first is by enhancing the efficacy of the well-established technique of Interplanetary Scintillations (IPS) by using the 32 simultaneous tied array beams. The second is by making technically challenging Faraday rotation measurements of background radio sources to determine the magnetic field content of Coronal Mass Ejections (CMEs) and explore their value in predicting space weather effects on Earth [6].

The wide field of view, high-quality PSF, high sensitivity, and the frequency and time resolution also make the MWA an ideal tool for transient studies [7]. The MWA will enable both blind and targeted source variability studies over a range of timescales from tens of nanoseconds to years. The design of the transient detection system is being optimized via the implementation of multiple specialized software analysis tools, operating both on the cross-correlation output data and on voltage-sum sample streams. There are four backend "instruments" planned for transient analyses, each of which is a combination of software applications and observing programs, and include an all-sky monitor, transient light-curve analyzers, and dedicated surveys of the whole accessible sky.

## III. CHALLENGES AT LOW FREQUENCIES

Any low frequency radio telescope design must confront three major technical challenges in order to enable comprehensive, high-fidelity measurements of the radio sky. These include: *(a)* Spatially and temporally variable ionospheric propagation effects, including refraction, refractive scintillation, ionospheric opacity, and Faraday rotation. These effects generally scale as $\lambda^2$; *(b)* Man-made radio frequency interference (RFI), of various kinds including ground-based, airborne and space-based, transmitted or incidental, propagated into the system by a variety of paths direct and indirect, from sources distant or local, including digital systems of the telescope itself; and *(c)* Wide-field calibration and imaging. Because it is physically impractical to build low frequency antennas with negligible response outside the sky region of interest, sky brightness distributions and instrumental response must be determined across very wide areas (generally the entire visible hemisphere). Hitherto these combined challenges have proved intractable, and low-frequency systems have significantly underperformed relative to their theoretical potential.

The MWA has tackled the RFI challenge primarily by selection of an extraordinarily radio-quiet site in outback Western Australia, sharply reducing the complexity and cost of the hardware (lower bit depth) and the extent and complexity of RFI mitigation strategies and algorithms that must be employed. The quality of the site has been validated through a series of field prototyping experiments conducted in 2005 [8]. These experiments revealed spectral occupancy at 1 kHz resolution across most of the band to be on the order of $10^{-4}$ in 15-second integrations, and less than $10^{-3}$ in several arbitrarily selected 4MHz bands with 30-60 minute integrations. In addition, a 10-hour integration across a 1 MHz band centered at 187 MHz showed no evidence of measurable RFI. The other two challenges (ionosphere and wide-field calibration and imaging) are approachable given sufficient, appropriate calibration information with which to simultaneously solve for sky structure, ionospheric structure, and instrumental response; details of the MWA approach and its feasibility can be found in [10]. Utilizing the information-maximization design approach described in the introduction above, the MWA will permit significantly improved solutions to these challenges, with accompanying enhancements in scientific capability over previous instruments.

## IV. HARDWARE DESIGN

The overall MWA hardware system design is illustrated in Fig. 1. The major subsystems of the array include the antenna tile and the analog beamformer that points the tile, the receiver node that digitizes the signal and selects the RF bandwidth for further processing, the correlator that produces the cross-correlations of all signal pairs, and the real-time computer which hosts the real-time software. In this section, we outline the system requirements and describe each of the subsystems, tracing the signal path from sky to calibrated image.

### A. System Requirements

In order to meet the key science objectives, the MWA must achieve specific performance goals, generally in terms of thermal noise limited sensitivity over various integration times and angular scales. The array must also have sufficient imaging quality and sensitivity to isolate and detect a large number of astronomical calibrator sources, both within an ionospheric calibration time interval (seconds) and an instrumental calibration timescale (minutes to hours). A list of the MWA specifications is given in Table 1.

### B. Antenna and Analog Beamformer

The antenna system is responsible for collecting radiation from the sky and presenting a suitably conditioned 80-300 MHz RF signal to the receiver node. A phased-array design for the antenna was selected, with a single, electronically steerable tile beam. (More than one tile beam would be possible, but is deemed not cost-effective.) The desired properties of the antenna include: *(a)* Zenith collecting area > 10 m$^2$ over as much of operating frequency range as possible; *(b)* Field of view from zenith to 60º ZA with <6 dB gain variation with beam pointing direction; *(c)* Minimal antenna gain at the horizon, to reduce terrestrial RFI; *(d)* System temperature dominated by sky noise; *(e)* full polarization response; and *(f)* Low manufacturing cost. In addition, the mechanical and electrical properties of the antenna system need to vary smoothly in time, frequency and angle, and variations from tile to tile need to be minimized, and manufacturing tolerances have been set accordingly, consistent with budgetary constraints.

The basic antenna element is a dual-polarization active dipole employing vertical bowtie elements that are symmetrical about the horizontal centerline, with an integrated LNA/balun at the juncture between the two arms of the bowtie. Compared with a simple, linear, horizontal dipole, the bowtie gives a broader antenna pattern and antenna impedance better matched to the LNA across the band. Constructing the bowtie in outline form, with spars around the outer edges and a single horizontal crosspiece, gives performance similar to a solid-panel bowtie at lower cost and weight.

Each LNA employs two Agilent ATF-54143 amplifiers in a balanced configuration with a balun on the output. The output signal from each LNA is carried via a 7-m-long coax cable to the analog beamformer. DC power for the LNA is carried on the same cable from the beamformer. Despite imperfect impedance matching between the bowtie and the LNA, the high levels of sky noise found at these low frequencies dominate the overall system temperature. Measurements indicate that the temperature of the sky in "cold" regions exceeds the contribution from the LNA and beamformer by a factor that reaches 3-5 over much of the frequency range, and approaches 1 at the lower and upper ends of the range. The galactic plane exceeds the temperature of "cold sky" by a factor of several.

An antenna "tile" consists of 16 dual-polarization dipoles arranged in a square 4x4 configuration with 1.1 meter center-to-center spacing as shown in Fig. 2. The spacing corresponds to $\lambda/2$ at 136 MHz, and was chosen to optimize the sensitivity in the frequency range thought likely to contain the EOR redshifted-HI signal. The dipoles are attached to a 5m x 5m area of galvanized steel mesh with 4mm wire and 5cm square openings, serving as a single ground plane for the tile. The mesh also serves as the registration grid for the dipole configuration, with tolerances across the tile on the order of 1cm. The choice of dipole spacing is a compromise between effective collecting area (and thus sensitivity) at low frequencies, versus grating lobes at high frequencies.

The analog beamformer receives dual polarization signals from all 16 crossed dipoles in a tile, and applies independent delays to each signal in a manner appropriate to form a tile beam in a particular direction on the sky. True delay steering is employed, rather than phase steering, in order to point the beam properly over the full operating frequency range. The delayed signals are combined, amplified, and sent over coaxial cable to the node receiver for digitization.

The beamformer employs five switchable delay lines for each dipole, each differing in delay by a factor of two, implemented as coplanar waveguide on four-layer, FR4 printed-circuit boards. The delays for all 32 signals (16 dipoles and 2 polarizations) may be set independently of each other via the monitor/control system. After the delay lines the 16 signals of the same polarization are combined in a passive combiner. Each of the two polarization signals then passes through a Walsh-function 180º phase switch and an amplifier before being carried to the receiver node over RG6 coaxial cable. The Walsh-functions provide orthogonality over the 16 signals handled by a single receiver, with a complete Walsh cycle every 160 msec. DC power at 48 volts, Walsh switching waveforms and beamformer switch settings for steering the tile, as well as communication from the beamformer back to the node receiver, are also carried on the same coaxial cable, yielding mechanical simplicity and low cable cost.

The delay line boards and control circuitry are housed in a steel beamformer enclosure that rests directly on the ground. Based on field experience, active cooling is not required. Also, no extraordinary measures are required to make the enclosure RFI-tight, since clocked signals (100 kHz) and associated weak emissions exist within the unit only during pointing changes, which last a few milliseconds every several minutes. Instead, the components and the delay line architecture have been designed for insensitivity to thermal changes, and for minimal RFI generation.

*C. Receiver Node*

A receiver node unit is responsible for digitizing, spectrally filtering and formatting data from 16 RF inputs (8 tiles, 2 polarizations), and transmitting the resulting digital data streams over fiber to the central processing location of the MWA ~2km away. There are 64 receiver nodes serving 512 tiles in the fully configured MWA system. Each node is physically located close to the 8 antenna tiles it serves, and is connected to the analog beamformers via RG6 coaxial cables. The node receiver also serves a variety of monitor and control functions, receives centrally-generated power, and distributes power to the tiles.

RF signals from the antenna elements are bandpass-filtered in each tile's beamformer. Upon arrival at the node, additional anti-aliasing filtering and low frequency rejection, signal level adjustments, and equalization (if necessitated by cable losses), are performed by an analog signal conditioning (ASC) board. The band-limited signals are fed in pairs to dual 8-bit analog-to-digital converter chips (ATMEL AT84AD001B) running at 655.36 Msample/sec. The dual outputs of each ADC chip are fed into a Xilinx Virtex 4 SX35 FPGA chip, in which is instantiated a dual 256-channel polyphase filterbank, yielding channel widths of 1.28 MHz. 24 of these channels, comprising 30.72 MHz of RF bandwidth, are then selected for further processing, formatted, and transmitted to the central facility via fiber. Data from all 16 input RF signals to the node receiver are formatted onto 3 fibers, each carrying data from all 8 tiles with both polarizations, but only one third of the frequency channels. Data are transmitted in the form of 5+5 bit complex samples.

Additional fibers provide Ethernet communications for monitor and control functions and distribute a centralized clock signal for the samplers, a clock for driving the FPGA logic, and timing signals for array synchronization. A single board computer controls the receiver node functions and services M&C needs. Control functions include analog beamformer commands, coordination of phase-switching signals to the beamformers, setting and monitoring of ASC units, configuring and monitoring of high-speed digital boards, and managing power startup and shutoff in response to various conditions.

The node receiver equipment occupies a volume of order 0.2 m$^3$, and dissipates up to 300 watts. It contains extensive high-speed digital circuitry capable of generating strong RFI, and must be placed in the field, close to the antenna tiles, in an environment where the ambient air temperature can exceed 50°C. The receiver unit thus requires a durable, weather-tight, cooled enclosure with a high degree of RF shielding, both in order to meet stringent radio quiet zone emissions regulations, and to avoid troublesome self-interference for the MWA.

The aggregate data rate for the MWA, with 30.72 MHz of processed bandwidth from 512 tiles, exceeds 300 Gbits/sec. Rapid cost reductions in fibers and transceiver hardware have made it easier to satisfy these data transport requirements.

*D. Digital Correlator and Array Beamformer*

Cross-correlation of all signal pairs in the MWA is a formidable computational challenge. There are 512 tiles times two polarizations, yielding 524,288 signal pairs to be multiplied together, each at 30.72 MHz bandwidth. This requires more than $1.6 \times 10^{13}$ complex multiply and accumulate (CMAC) operations per second, in an environment where power consumption and associated cooling are major practical and economic concerns. Furthermore, the array architecture requires that computationally expensive spectral filtering (to 10 kHz spectral resolution) precede the multiplication step.

The correlator subsystem performs two main tasks, spectral filtering and cross-multiplication, that are implemented on two different FPGA-based board types, referred to as the polyphase filterbank board (PFB) and the correlator board (CBD). The full system comprises 16 PFBs and 72 CBDs, mounted in ATCA card cages, along with a slot-1 controller board, a full-mesh backplane, and small physical interface boards, known as rear transition modules (RTMs). The full system occupies 8 card cages in 3 standard size racks, and the power consumption is ~15 kW.

The system receives 192 data fibers from 64 receiver nodes, carrying 24 spectral channels of 1.28 MHz each from all 512 dual-polarization tiles, for a total of 24,576 data streams each 1.28 MHz wide. The PFB boards execute a 128-channel operation on each of these streams, yielding a spectral resolution of 10 kHz, and a set of 3,145,728 data streams. These streams are re-ordered to facilitate the cross-multiply operation, and transmitted to the CBDs over the backplane.

The cross-multiply and accumulate operation is done on the 10 kHz data streams, in FPGA devices which have typical clock speeds on the order of 250 MHz, some 25,000 times faster than the incoming sample rate. The design therefore employs extensive data buffering and multiplexing in time and frequency in order to service many data streams, forming many baselines at many frequencies in each FPGA multiplier unit, within a single data stream sample interval. By placing 8 SX35 FPGA chips on each board, each employing 132 multiplier units, and using 72 boards for a total of 576 chips and 76,032 multipliers, the necessary 1.6 billion complex cross products are formed within the 100 μsec sample interval of the data streams. These cross products are accumulated for 0.5 seconds before being packaged and transmitted over gigabit Ethernet to the real time computer, which is physically co-located with the correlator. The architecture allows for flexible reconfiguration to accommodate different levels of efficiency in the final FPGA correlation code, impacting total hardware volume by no more than 15%.

Traditional radio astronomy correlators for imaging arrays are designed to compensate for changing array geometry and Doppler shifts due to earth rotation, "stopping" the fringes in hardware and removing geometrical delays in order to permit coherent integration of the correlated signals in time and frequency. For the MWA, however, a different approach has been adopted. Due to the physically small array extent and long wavelengths, it is practical to handle the correlated data in small enough segments of time and frequency that the phase differences across the segments are small, and decorrelation is negligible. With this approach, no geometrical calculations are required, and the hardware correlator is dramatically reduced in complexity. The phase pointing center of the correlator, conventionally set to track a specific RA and DEC, is instead fixed at the zenith. Essentially, the correlator is thus hardwired to a single mode, and requires only basic real time control software. The principal advantage of this arrangement is that it transfers complexity away from correlator hardware, firmware and online software, and into general purpose computing hardware and software where such complexity is easier to deal with. The downside is that the correlator output data volume is high by conventional standards.

*E. Real Time Computer and Data Storage*

Data flows from the correlator to the real-time computer (RTC) at up to 160 Gbits/sec, over gigabit Ethernet connections. For the initial implementation, additional frequency averaging will be implemented in the correlator hardware, reducing this rate to 40 Gbits/sec at the cost of some flexibility. These data must be processed in order to generate high-precision ionospheric and instrumental calibration solutions, with new solutions, and images generated using those solutions, every 8 seconds. The required computing rate under normal conditions is estimated to be between 2 and 10 Tflop/s, with greater accuracy and greater tolerance of adverse observing conditions toward the upper end of this range. The nature of the calculations has been shown to be well-matched to the capabilities of graphical processing units (GPUs), and the RTC will be implemented as a GPU-enhanced cluster with a 2-rack footprint and power consumption of ~20 kW.

V. ARRAY CONFIGURATION

The MWA array configuration is an interesting optimization exercise. The overall size of the array is determined by calibration issues and science requirements. Optimization of the array configuration is then driven by imaging characteristics, cabling costs for connecting the array components, and considerations of local topography and disallowed locations described by a site mask. The array configuration problem lies in the class of *NP complete* problems for which determining the formally optimal solution is impractical, and we have adopted a heuristic approach to determining an acceptable solution. The large number of elements needing to be distributed over a comparatively limited area, leading to an essentially Nyquist sampled u-v space, eliminate the need for fine tuning the placement of individual elements.

The MWA collecting area is distributed pseudo randomly in a heavily centrally condensed manner. The full array configuration, along with the cable network, is shown in Fig. 3 which includes a zoomed view into the central region. The u-v coverage and resultant PSF corresponding to the array configuration are shown in Figs. 4 and 5. The high density of the u-v coverage and the excellent PSF quality are evident.

Of the 512 MWA tiles, 496 are placed within a circular region 1.5 km in diameter. The remaining 16 tiles are distributed over a roughly 3 km region. The tiles are distributed with a uniform areal density in the central ~50m region of the array with an inverse square fall off in the areal density in the region beyond that. A site mask which includes topological features and vegetation information, among other things, was used to constrain permitted locations for tiles. It is worth noting that the chosen site of the MWA is exceptionally flat, with the heights of all the tiles in the 1.5 km diameter being within 6m of one another. This allows the full 3-dimensional nature of the array to be accommodated with a very small computational impact.

## VI. SOFTWARE DESIGN

There are two main components of the software for MWA. These are the real time calibration and imaging system (referred to as the "real-time system" or RTS) and the monitor and control system.

### A. Real Time System (RTS)

This software system is responsible for processing the raw visibility data from the correlator, determining both instrumental and ionospheric calibration parameters, applying the calibrations, and generating images. This processing must occur in real time, since the raw data throughput of the correlator is too large for storage and subsequent offline analysis. In other words, since there are too many visibilities to store, the data must be reduced to images in real-time, a generalized self-calibration must be performed in real-time, and averaging must be performed in image space. The RTS cadence was chosen to be 8 seconds which has been found to be an optimal choice based on considerations of array sensitivity, ionospheric variability time scales, source count distributions and confusion limits for the array. The calibration system has been described in detail by Mitchell et al. [10].

*1) Ionospheric Calibration*

At the low frequencies of interest for MWA, the field of view at ionospheric heights (100-1000 km) is large compared to the characteristic scale size of the ionospheric irregularities responsible for phase distortions. This means that position-dependent solutions for the ionosphere are required, and that robust information from calibration sources across the field of view be gathered.

In order to simplify the ionospheric calibration problem, the MWA design limits the maximum baseline to 3 km, which is typically much smaller than the ionospheric irregularity scale. Ionospheric phase effects across the array are thus well approximated by a linear phase ramp, producing a refractive shift in the apparent position of a source.

The refractive shift, which varies across the field of view, is measured for each independent calibration source, yielding data for a model of the overall refractive distortion field. The effectiveness of the model depends on the areal density of calibration sources, the source position measurement accuracy, and the rate at which the model can be updated, to keep up with rapidly changing ionospheric conditions. This means that the array must be sufficiently sensitive to detect a dense grid of sources within a few seconds, and that PSF sidelobe confusion does not compromise the position measurements. The MWA will observe hundreds of point sources in its primary beam each RTS cycle. Strong sources outside the primary beam must be individually subtracted for accurate measurements on the fainter sources.

We note that at times of significant ionospheric scintillation, calibration quality will be compromised, and the range of useful observations that are possible will be reduced. The duty cycle of such adverse conditions is expected to be low at this mid-geomagnetic latitude site.

*2) Instrumental calibration*

A major challenge for the MWA is achieving a high-precision calibration of the full polarization response of the instrument. Without this, the wide field of view, the crowded and complex polarized sky, and the modulation of calibration errors by the time-variable ionosphere will render the primary scientific goals unachievable.

The instrumental response is determined by the responses of each of the 512 individual antenna tiles, as a function of direction on the sky, observing frequency, and time. While in practice there are both direction-independent and direction-dependent components of the complex gain of a given tile, in this discussion we will treat the former as a special case of the latter. The essential problem is that the RF output from a single tile represents the sum of signals from the whole sky, modified by the tile response across the sky, and it is not possible to isolate contributions from different directions. Thus, a mechanism is needed to isolate signals from individual calibration sources on the sky so that the instrumental complex gain in a specific direction can be measured. The MWA design allows this to be done effectively, due to the large number of antennas and the correspondingly low sidelobes of the phased-array beam [10].

Conceptually, the MWA uses an iterative algorithm in which 511 of the 512 antenna tiles are phased up in the direction of a calibration source, so that the resulting voltage sum is dominated by the signal from the source with good rejection of the rest of the sky. This voltage sum can then be cross-correlated with the $512^{th}$ tile, yielding information on the complex response of that tile in the direction of the source. Cycling through all antennas then improves the quality of the phasing step, and the process can be convergent. The quality of the convergence strongly depends on the degree to which the phasing operation isolates the target source signal from the rest of the sky. This isolation is excellent for the 512-tile MWA design, so the algorithm would be expected to be efficient. The process is then repeated for all calibrators and frequencies, thereby building up a model of the complex direction-dependent gain for each tile. This is an extended and customized variant of the SUMPLE algorithm [9].

Such an algorithm, if implemented by voltage sum beamforming in the manner described, would be prohibitively expensive for the MWA. However, it is instead possible to compute the complex visibility that would be formed by cross-correlating the voltage sum of 511 tiles with the $512^{th}$ tile merely by direction-dependent coherent addition of the visibilities produced by the correlator. In this implementation, the correlator does most of the calculations associated with "phasing up" each set of 511 antenna tiles, leaving the RTS a much less computationally demanding task of operations on averaged visibilities. The algorithm also adds an essential step to further suppress unwanted responses from directions other than the target source. The brightest sources in the sky, even outside the primary beam, are targeted first, and are then subtracted from the data set in order to eliminate their signals when phasing up on weaker sources. Implementation details and performance estimates are given in [10].

The implementation is designed to produce a full polarization calibration solution. The measurements made towards each calibrator include the response of each instrument polarization to each source polarization (fits are of Jones matrices, as in [10]). Most of the calibrators are unpolarized, and will be used to constrain most of the degrees of freedom in the primary beam models. Polarized sources such as pulsars will be used to constrain the remaining degrees of freedom.

*3) Imaging*

The MWA is unlike traditional large dish radio telescopes in that the antenna tiles will have primary beams with power patterns that, due to manufacturing and installation tolerances, cannot be considered to be identical. In addition, MWA will suffer from large instrumental polarization due to the changing projection of the element dipoles on the sky. The purpose of the imaging pipeline is to produce calibrated snapshot images of the sky within the primary beam each RTS cycle in Stokes parameters and in a celestial coordinate frame. The imaging system is required to correct for instrument calibration, ionospheric refraction and Faraday rotation.

The pipeline consists of four basic steps for each RTS cycle: (i) gridding the visibility data onto a regular coordinate grid; (ii) transforming into image space to generate four snapshot images of the ionospherically-distorted sky in instrument polarization; (iii) converting the images from instrumental polarization to Stokes parameters using the current calibration solutions for the instrument and ionosphere for Faraday rotation; (iv) re-gridding the images, where the ionosphere-distorted sky is interpolated into a common celestial coordinate frame.

Corrections for the ionosphere and instrument polarization are performed in image space. Consequently, the calibration solution must be applied on a baseline by baseline basis to maintain good polarization purity and to prevent calibration errors from introducing artifacts into the snapshots. Calibration solutions are applied during the gridding process [11]. The image space processes to correct the ionosphere and convert to Stokes are highly parallel pixel-based operations and ideally suited for GPUs.

Each snapshot must be re-gridded to remove the ionospheric distortion and the choice of projection is arbitrary. As the instrument will be capable of imaging the entire sky south of declination +40°, a projection that accommodates a large fraction of the celestial sphere with minimal distortions is preferred. To this end we have chosen HEALPIX [12, 13] as our output pixelization. Every snapshot is re-gridded into this pixelization and stored internally as an ordered list of HEALPIX pixels. The algorithm that is used to re-project the data from a distorted instrument frame into the HEALPIX pixelization must do so rapidly, but also introduce minimal artifacts. Several methods are being evaluated.

*B. Monitor and Control*

The Monitor and Control (M&C) subsystem is responsible for controlling the behavior of the end-to-end system from antennas to real-time data processing. It also receives, processes, and logs monitoring information of various kinds, including environmental data, subsystem performance metrics, and calibration data. Finally, it provides a user interface for controlling the instrument.

The M&C system is built around the meta-data archive which describes the MWA as a state machine. All the detailed information needed to completely specify the state of MWA is stored in the meta-data archive. It serves as an archive of this information for times in the past, defines the present state of the instrument, and serves as the scheduler for the array for times in the future. All the information gathered by the monitor part of the M&C system is also stored in the meta-data archive as is other relevant information from multiple sources e.g. RFI monitors, GPS systems and ionospheric and instrumental calibration solutions from the RTS.

The *monitor* part of the M&C system involves monitoring both the state of health of the MWA hardware from the antenna tiles to the RTC and the environmental conditions. The former includes monitoring parameters such as voltage and current draws through the array and acknowledgements of the commands issued to various sub-systems, and the latter includes monitoring of sensors such as weather, rack temperatures and generator fuel levels. The *control* part of the M&C system reads the MWA state information from the meta-data archive and issues the commands to various sub-systems to configure them. It also deals with the remote operations aspects of the array including powering the system on and off. The M&C system also provides a web based GUI interface to query the database and to define and schedule observations.

VII. CONCLUSION

The Murchison Widefield Array embodies a number of innovations and represents the most aggressive move to date towards the future-oriented concept of full sampling of the electric field across the telescope aperture, coupled with complete algorithmic control of the associated digital voltage data. This has been made possible by the current availability of powerful high-speed computational and digital signal processing systems. The resulting array capabilities are revolutionary, with high fidelity imaging at high spectral resolution over a wide fractional bandwidth and over a very wide field of view of many hundreds of square degrees. The survey speed of the array, per unit hardware cost, is extraordinary and provides rich scientific opportunities.

**Table 1. MWA Specifications**

| | |
|---|---|
| Frequency range | 80-300 MHz |
| Number of receptors | 8192 dual polarization dipoles |
| Number of antenna tiles | 512 |
| Number of baselines | 130,816 |
| Collecting area | ~8000 $m^2$ (at 200 MHz) |
| Field of View | ~15°-50° (1000 $deg^2$ at 200 MHz) |
| Configuration | Core array ~1.5 km diameter (97% of area) + extended array ~3 km diameter (3% of area) |
| Bandwidth | 220 MHz (Sampled); 30.72 MHz (Processed) |
| Spectral channels | 1024 (30 kHz spectral resolution) |
| Temporal resolution | 0.5 sec uncalibrated, 8 sec calibrated |
| Polarization | Full Stokes |
| Continuum point source sensitivity | 20mJy in 1 sec (at 200 MHz full bandwidth); 0.34mJy in 1 hr |
| Array voltage-beams | 32, single polarization |

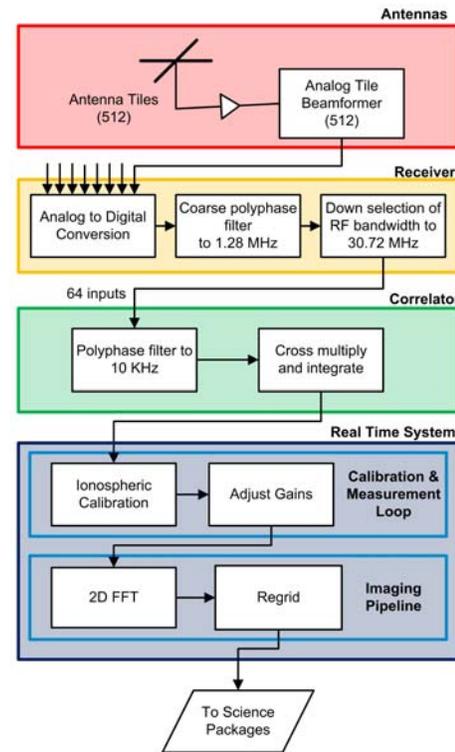

*Fig.1. Simplified block diagram of the MWA hardware system.*

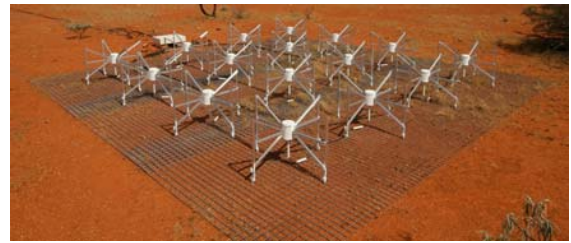

*Fig.2. Photograph of one of the 32 MWA tiles currently deployed at Boolardy, Western Australia. The 5 x 5 m wire mesh ground plane is laid directly on the ground and the dipoles are clipped onto the mesh. The beamformer box is seen in the background.*

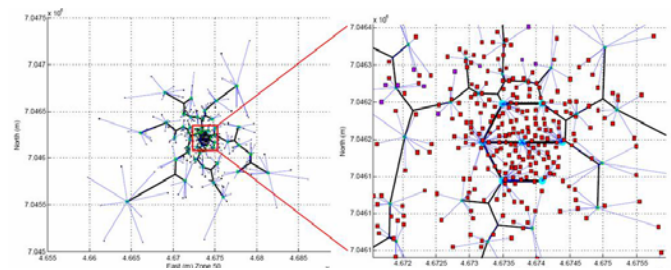

*Fig. 3. The left panel shows the entire array configuration with a grid spacing of 500m. Red squares represent tiles, and thin blue and thick black lines represent co-axial cables connecting the tiles to beamformers and the fiber optic cable network, respectively. The right panel shows a zoomed view of the central part of the configuration with a grid spacing of 50 m.*

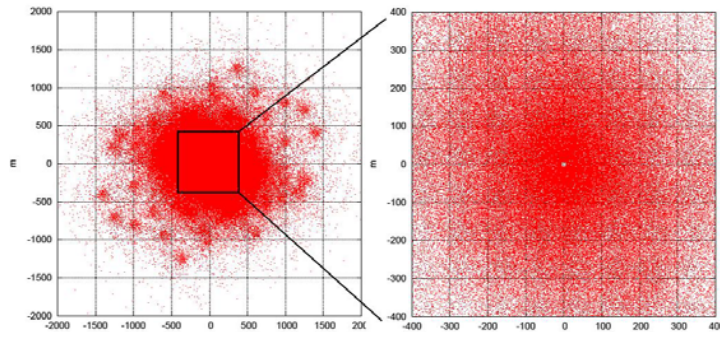

Fig. 4 The left panel shows the instantaneous monochromatic uv coverage for the case of zenith pointing and λ=1m. The right panel shows a zoomed-in view of the central part of the uv plane as indicated.

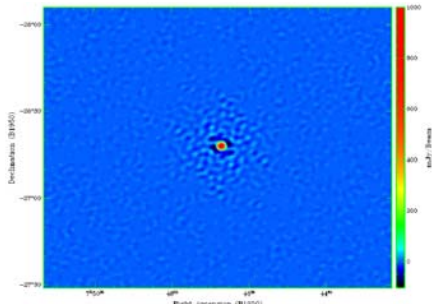

Fig. 5. The normalized PSF for the MWA corresponding to the uv coverage shown in Fig 4, with the array pointed towards zenith. The brighter features in the sidelobe structure seen close to the central peak are at about 3-4% level and the ones far out are ~0.3% level. The deepest minimum is -10.2%.